\documentclass[aps,prd,amsfonts,amssymb,amsmath,12pt,%
  notitlepage,nofootinbib,groupedaddress]{revtex4-1}
\usepackage[dvips]{graphicx}
\usepackage{bm}

\begin{document}
\title{Quark-antiquark potential to order \bm{$1/m$}\\ and heavy quark masses}
\author{Alexander Laschka}
\author{Norbert Kaiser}
\author{Wolfram Weise}
\affiliation{Physik Department, Technische Universit\"{a}t M\"{u}nchen,
  D-85747 Garching, Germany}
\date{May 3, 2011}

\begin{abstract}
An updated heavy quark-antiquark potential is constructed by matching the
short-distance perturbative part to long-distance lattice QCD results at an
intermediate $r$ scale. The static potential and the order $1/m$ potential are
both analyzed in this way. Effects of order $1/m$ in charmonium and bottomonium
spectra are discussed in comparison. Charm and bottom quark masses are deduced
from the spectra and related to the quark masses of other schemes.
\end{abstract}

\maketitle

\section{Introduction}
Investigations of the potential between a heavy quark and its antiquark have a
long tradition. Early models featuring the basic Coulomb-plus-linear
$r$ dependence plus a hyperfine interaction established a quite successful
phenomenology of charmonium spectroscopy below the open charm threshold. At
short distances perturbative QCD is supposed to work. At larger distances
lattice QCD results have confirmed the linearly rising confinement potential
and established reliable values of the string tension.

It has long been realized that the static quarkonium potential, constructed
within the framework of QCD perturbation theory, has a badly convergent or even
divergent behavior at short distances. This has been understood in terms of
renormalon ambiguities. It has in fact been found that leading renormalon
effects cancel in the sum of the static potential and twice the quark (pole)
mass~\cite{Beneke:1998rk,Hoang:1998nz}, an important feature to be recalled
later when we address the issue of quark masses in the context of the
quarkonium potential and spectroscopy.

The steadily increasing precision of current lattice QCD computations permits
now an accurate matching of perturbative and non-perturbative approaches at an
intermediate distance scale. An update of this procedure, combined with an
accurate analysis of charm and bottom quark masses, is the main topic of the
present work.

This paper is organized as follows. After a brief summary of first attempts in
Section~\ref{sec:first-attempts}, the perturbative static potential in
coordinate space and its overall additive constant are derived in
Section~\ref{sec:static-potential} and matched at an appropriate distance scale
with results from lattice QCD. Section~\ref{sec:order-1/m-potential} follows
with an analogous construction for the potential at order $1/m$ in the quark
mass. Section~\ref{sec:quarkonium-spectra} discusses $1/m$ effects on
bottomonium and charmonium spectra. In the last section charm and bottom quark
masses are deduced and compared with existing determinations in other schemes.

\section{Preparations}
\label{sec:first-attempts}
As a starting point, consider the static quark-antiquark potential in momentum
space at three-loop order:
\begin{multline}
  \label{eq:static_potential_1}
  \tilde{V}^{(0)}(|\vec q\,|) =-\frac{4\pi C_F
  \alpha_s(|\vec q\,|)}{\vec q\,^2}\,
  \bigg\{1+\frac{\alpha_s(|\vec q\,|)}{4\pi}\, a_1
  +\left(\frac{\alpha_s(|\vec q\,|)}{4 \pi}\right)^2 a_2\\
  +\left(\frac{\alpha_s(|\vec q\,|)}{4 \pi}\right)^3
  \left(a_3+8\pi^2 C_A^3 \ln\frac{\mu_{\scriptscriptstyle\text{IR}}^2}%
  {\vec q\,^2}\right) +\mathcal O \left(\alpha_s^4\right) \bigg\}\, ,
\end{multline}
where $\vec q$ is the three-momentum transfer.
The coefficients $a_1$ and $a_2$ have been determined analytically and they
read in the $\overline{\text{MS}}$
scheme~\cite{Peter:1996ig,Peter:1997me,Schroder:1998vy}:
\begin{align}
  a_1 &= \frac{31}{9}C_A - \frac{20}{9}T_F n_f,\\
  a_2 &= \left( \frac{4343}{162} + 4\pi^2 - \frac{\pi^4}{4}
    + \frac{22}{3}\zeta(3)\right)C_A^2
    - \left( \frac{1798}{81}
    +\frac{56}{3}\zeta(3)\right)C_A T_F n_f\notag\\
    &\hspace{2em} -\left(\frac{55}{3}-16\zeta(3)\right)C_F T_F n_f
    +\left( \frac{20}{9}T_F n_f\right)^2,
\end{align}
where $C_F=4/3$, $C_A=3$, $T_F=1/2$ for SU(3) and $n_f$ is the number of
light quark flavors. At three-loop order, infrared singular contributions
$\ln(\mu_{\scriptscriptstyle\text{IR}}^2/\vec q\,^2)$ start to play a role
(see e.g.~\cite{Brambilla:1999qa}). The accompanying constant
\begin{equation}
a_3=4^3\,(209.884(1)-51.4048n_f+2.9061n_f^2-0.0214n_f^3)
\end{equation}
has been calculated independently in~\cite{Smirnov:2009fh}
and~\cite{Anzai:2009tm}. In this paper we focus on the two-loop level. In order
to transform the potential to coordinate space, $\alpha_s(|\vec q\,|)$
in Eq.~(\ref{eq:static_potential_1}) is usually expressed (see
e.g.~\cite{Kniehl:2002br}) as a powers series expansion in $\alpha_s$ at some
fixed scale $\mu$:
\begin{align}
  \label{eq:power_exp}
  \alpha_s(q) = \alpha_s(\mu)
   \bigg[1&-\frac{\alpha_s(\mu)}{4\pi}\beta_0\,\ell
  + \left(\frac{\alpha_s(\mu)}{4\pi}\right)^2 (\beta_0^2\,\ell\!
  -\!\beta_1)\,\ell\notag\\
  &+ \left(\frac{\alpha_s(\mu)}{4\pi}\right)^3
  \left(-\beta_0^3\,\ell^2\! +\! \frac{5}{2}\beta_0\beta_1\,\ell\!
  -\!\beta_2 \right)\ell\notag\\
  &+ \left(\frac{\alpha_s(\mu)}{4\pi}\right)^4
  \bigg(\beta_0^4\,\ell^3\! -\! \frac{13}{3}\beta_0^2\beta_1\,\ell^2
  \!+\! 3\Big(\beta_0 \beta_2\! +\! \frac{\beta_1^2}{2}\Big)\ell\!
  -\! \beta_3 \bigg)\ell
  +\mathcal O \left(\alpha_s^5\right) \bigg],
\end{align}
with $\ell=\ln(q^2/\mu^2)$. The coefficients $\beta_n$ of the QCD
$\beta$ function are known up to four-loop order~\cite{vanRitbergen:1997va}:
\begin{equation}
  \label{eq:beta_function}
  \frac{\partial\alpha_s (\mu)}{\partial\ln \mu^2} =\beta(\alpha_s)
  = -\frac{\alpha_s^2}{4\pi}\beta_0
    -\frac{\alpha_s^3}{(4\pi)^2}\beta_1-\frac{\alpha_s^4}{(4\pi)^3}\beta_2
    -\frac{\alpha_s^5}{(4\pi)^4}\beta_3+\mathcal O(\alpha_s^6).
\end{equation}
In this approach a Fourier transform leads to the standard, $\mu$-dependent
definition of the coordinate space static potential,
\begin{multline}
  \label{eq:static_potential_2}
  V^{(0)}(r) = - \frac{4\,\alpha_s(\mu)}{3\,r}\, \bigg\{1 
  + \frac{\alpha_s(\mu)}{4\pi}\,\Big[ a_1 + 2\beta_0\, g_\mu(r) \Big]\\
  + \bigg(\frac{\alpha_s(\mu)}{4\pi}\bigg)^2 \Big[ a_2 
  + \beta_0^2\big(4g_\mu^2(r)\! +\! \pi^2/3\big)
  + 2g_\mu(r)(2a_1\beta_0\!+\!\beta_1)\Big]\\
  +\bigg(\frac{\alpha_s(\mu)}{4\pi}\bigg)^3 \Big[ a_3
  +16\pi^2 C_A^3\big( \ln (\mu_{\scriptscriptstyle\text{IR}} r)\!
  +\!\gamma_{\scriptscriptstyle\text{E}}\big)
  +\beta_0^3\big(8g_\mu^3(r)\!+\!2\pi^2 g_\mu(r)\!+\!16\zeta(3)\big)\\
  +\beta_0\big(12g_\mu^2(r)\!+\!\pi^2\big)
  \big(a_1\beta_0\!+\!5/6\,\beta_1\big)
  +2g_\mu(r)\big(3a_2 \beta_0\!+\!2a_1\beta_1\!+\!\beta_2\big)\Big]
  +\mathcal O(\alpha_s^4) \bigg\} \, ,
\end{multline}
where $g_\mu(r)=\ln(\mu r)+ \gamma_{\scriptscriptstyle\text{E}}\,$. The
derivation of this $r$-space potential uses, in principle, information about
$\alpha_s(|\vec q\,|)$ over the full range in $q$ space.
\begin{figure}[tp]
{\centering
\includegraphics[width=.70\textwidth]{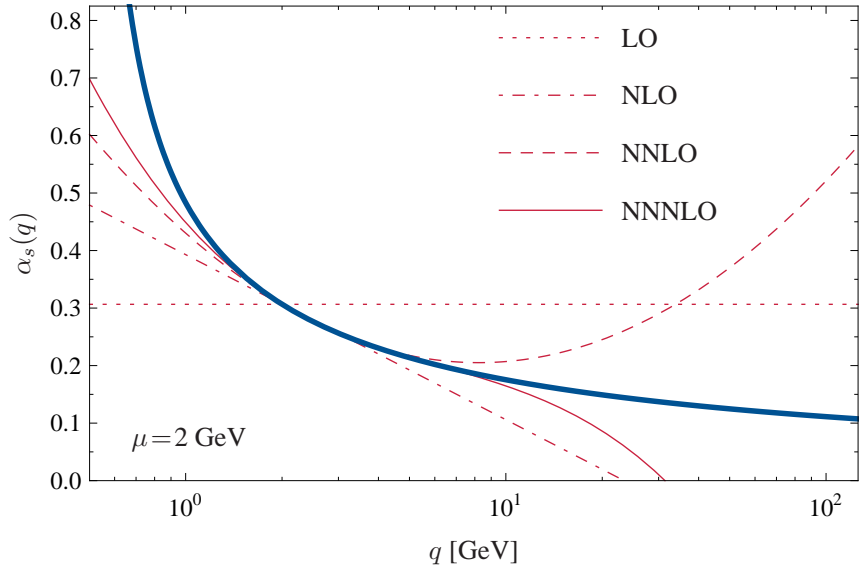}\\
}
\caption{\label{fig:1}Power series expansion of $\alpha_s(q)$
   at $\mu\!=\!2$~GeV for $n_f\!=\!4$. The thick line shows the full four-loop
   running according to the renormalization group equation. Note that the sign
   of $\alpha_s(q\to \infty)$ changes order by order.}
\end{figure}
However, the expansion~(\ref{eq:power_exp}) in powers of $\ln q^2$ is a good
approximation only in a small neighborhood of the scale $\mu$, as demonstrated
in Fig.~\ref{fig:1} for the choice $\mu=2$~GeV. Clearly, the behavior of
$\alpha_s(q)$ for $q>10$~GeV and $q<1$~GeV is out of control for such an
expansion.

\begin{figure}[tp]
{\centering
\includegraphics[width=0.70\textwidth]{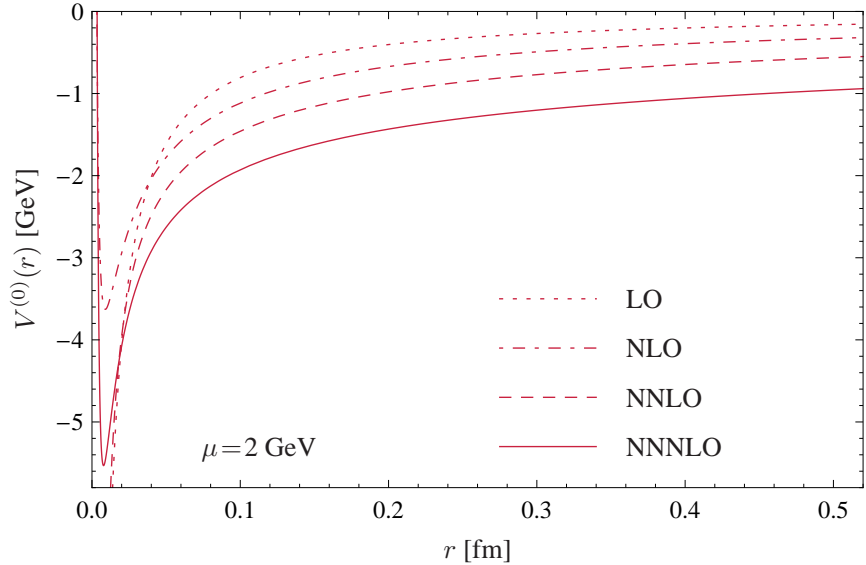}\\
}
\caption{\label{fig:2}Static $r$-space potential according to
  Eq.~(\ref{eq:static_potential_2}) using $\mu\!=\!2$~GeV. The choice
  $\mu_{\scriptscriptstyle\text{IR}}^2\!=\!\vec q\,^2$ has been adopted at
  NNNLO.}
\end{figure}
Fixing $\mu$ for instance at 2~GeV, the resulting coordinate space
potential~(\ref{eq:static_potential_2}) behaves pathologically at $r\!\to\! 0$,
as shown in Fig.~\ref{fig:2}. This behavior can be traced to the order-by-order
sign changes observed in Fig.~\ref{fig:1} for $q\!\to\!\infty$. It can be
improved using the renormalon subtracted scheme
(see e.g.~\cite{Pineda:2002se,Brambilla:2009bi}).
\begin{figure}[tp]
{\centering
\includegraphics[width=.70\textwidth]{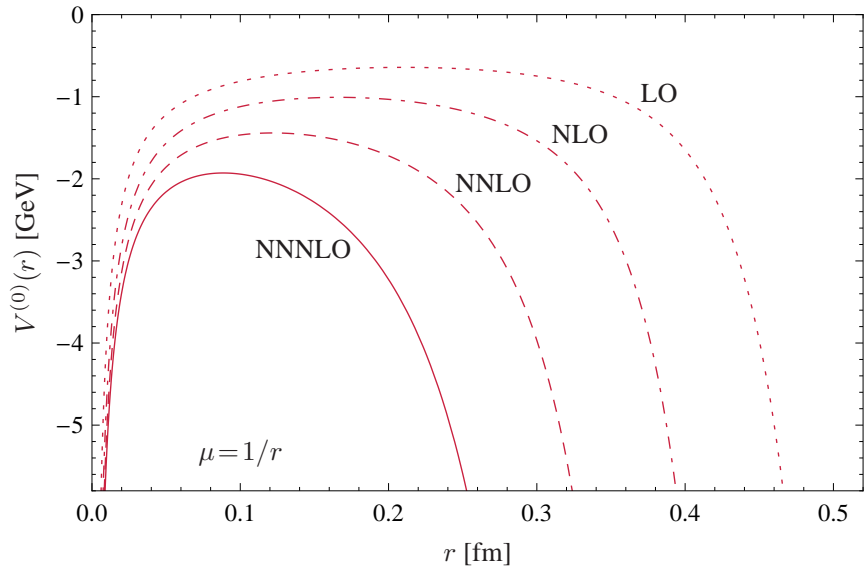}\\
}
\caption{\label{fig:3}Static $r$-space potential according to
  Eq.~(\ref{eq:static_potential_2}). Progressive orders are shown when $\mu$ is
  identified with $1/r$. The choice
  $\mu_{\scriptscriptstyle\text{IR}}^2\!=\!\vec q\,^2$ has been adopted at NNNLO.}
\end{figure}
Figure~\ref{fig:3} shows the potential resulting from the frequently used
\emph{ad hoc} identification $\mu=1/r$ which evidently works only at extremely
short distances, $r<0.02$~fm. Such a construction therefore does not suggest
itself for a matching of the potential to lattice QCD results at typical
distances $r\gtrsim 0.1$~fm.

\section{The static potential}
\label{sec:static-potential}
Here we pursue a different strategy for constructing the static potential in
coordinate space, based on the potential-subtracted (PS) scheme proposed by
Beneke~\cite{Beneke:1998rk}. The $r$-space potential is defined through a
restricted Fourier transform as
\begin{equation}
  \label{eq:r_space_potential}
  V^{(0)}(r,\mu_f) = \intop_{|\vec q\,|>\mu_f}\!\!
  \frac{d^3q}{(2\pi )^3}\ e^{i\vec q\cdot\vec r}\,
  \tilde{V}^{(0)}(|\vec q\,|)\, ,
\end{equation}
where $\tilde{V}^{(0)}(|\vec q\,|)$ is given in
Eq.~(\ref{eq:static_potential_1}), but now $\alpha_s(|\vec q\,|)$ for
$|\vec q\,|>\mu_f$ is used without resorting to a power series expansion.
The momentum space cutoff $\mu_f$ is introduced in order to delineate the
uncontrolled low-$q$ region from the high-$q$ range where perturbation theory
is considered to be reliable. The potential~(\ref{eq:r_space_potential})
differs from the ``true'' static potential,
\begin{equation}
  V^{(0)}(r) = v^{(0)}(\mu_f)+V^{(0)}(r,\mu_f)\, ,
\end{equation}
approximately by a constant,
\begin{equation}
  v^{(0)}(\mu_f) = \intop_{|\vec q\,|\leq\mu_f}\!\!
  \frac{d^3q}{(2\pi )^3}\ e^{i\vec q\cdot\vec r}\,
  \tilde{V}^{(0)}(|\vec q\,|)
  = \frac{1}{2\pi^2}\intop_0^{\mu_f}\!
  dq\,q^2\,
  \tilde{V}^{(0)}(q) + \mathcal{O}(\mu_f^2\, r^2)\, ,
\end{equation}
which encodes non-perturbative low-$q$ behavior that can be absorbed in the
definition of the potential-subtracted (PS) quark mass
(see Section~\ref{sec:quark-masses}). The correction of order $\mu_f^2\, r^2$
is negligibly small in the range of interest ($r<0.1$~fm).

\begin{figure}[tp]
{\centering
  \includegraphics[width=.70\textwidth]{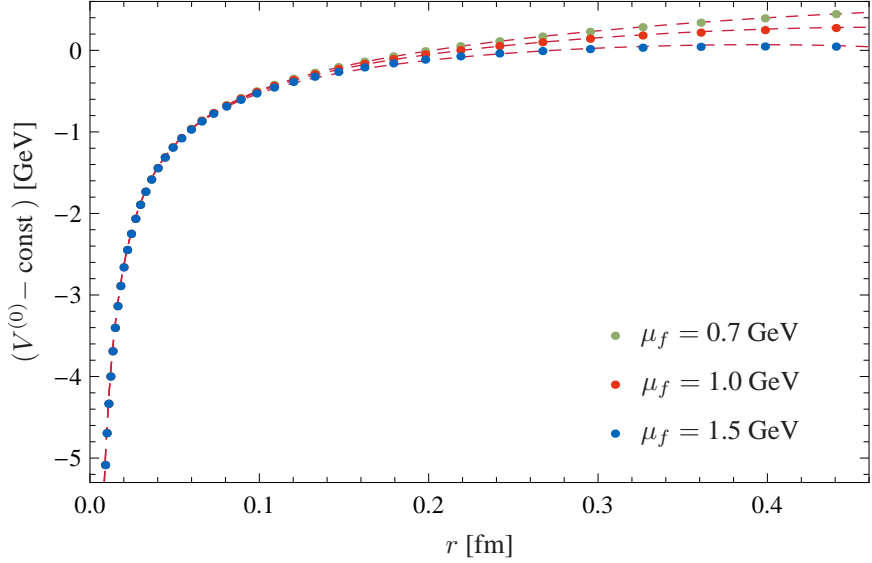}\\
}
\caption{\label{fig:4}Static QCD potential (with $n_f=3$) from the restricted
  numerical Fourier transform~(\ref{eq:r_space_potential}). Shown is the NNLO
  potential for different values of $\mu_f$. The curves have been shifted by
  a constant to match at small $r$ values.}
\end{figure}
\begin{figure}[tp]
{\centering
  \includegraphics[width=.70\textwidth]{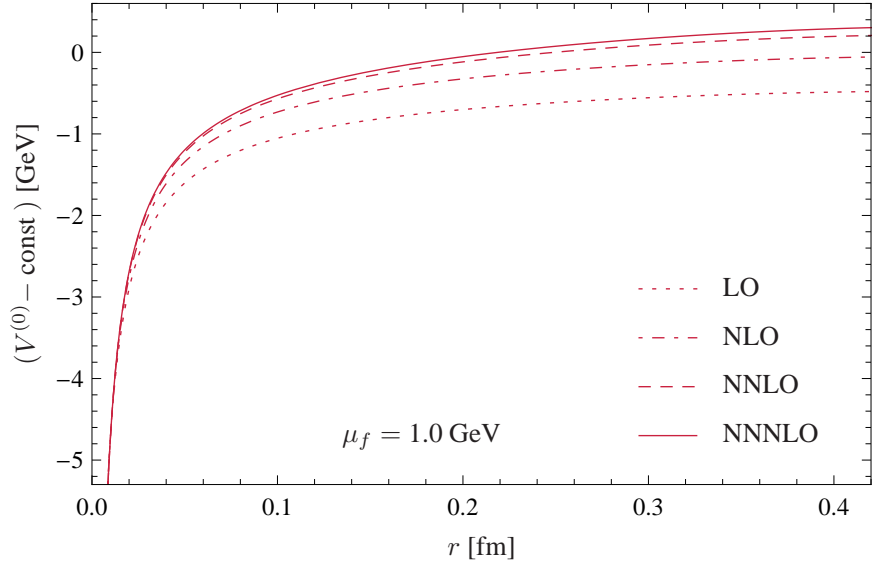}\\
}
\caption{\label{fig:5}Static QCD potential (with $n_f=3$ and
  $\mu_f\!=\!1.0$~GeV) from the restricted numerical Fourier
  transform~(\ref{eq:r_space_potential}). Different orders of have been matched
  at $0.01$~fm. The choice $\mu_{\scriptscriptstyle\text{IR}}^2\!=\!\vec q\,^2$
  has been adopted at NNNLO.}
\end{figure}
The potential $V^{(0)}(r,\mu_f)$ is evaluated numerically using the four-loop
renormalization group running of the strong coupling $\alpha_s$, see
Eq.~(\ref{eq:beta_function}). For distances $r < 0.2$~fm, the resulting
potential depends only marginally on $\mu_f$ as shown in Fig.~\ref{fig:4}. At
the matching radius, $r=0.14$~fm, the spread of $(V^{(0)}-\text{const})$ when
varying $\mu_f$ between $0.7$~GeV and $1.5$~GeV is $0.05$~GeV. The convergence
behavior of the potential is displayed in Fig.~\ref{fig:5}. Different orders
have been matched at $r=0.01$~fm and are then evolved to larger distances.
Evidently, the convergence behavior of the potential $V^{(0)}(r,\mu_f)$ is
satisfactory.

For bottomonium ($n_f\!=\!4$ massless flavors), the input value for the
renormalization group running of the strong coupling constant is chosen as
$\alpha_s(4.2\ \text{GeV})=0.226\pm 0.003$. In the case of charmonium
($n_f\!=\!3$) we use $\alpha_s(1.25\ \text{GeV})=0.406\pm0.010$ as input in the
potential. These values are obtained from
$\alpha_s(m_Z\!=\!91.1876\ \text{GeV})=0.1184\pm 0.0007$%
~\cite{Bethke:2009jm} (for a theory with $n_f\!=\!5$ active quark flavors) and
run down to $4.2$~GeV and $1.25$~GeV, taking into account flavor
thresholds~\cite{Chetyrkin:1997sg}.

\begin{figure}[tp]
{\centering
  \includegraphics[width=.70\textwidth]{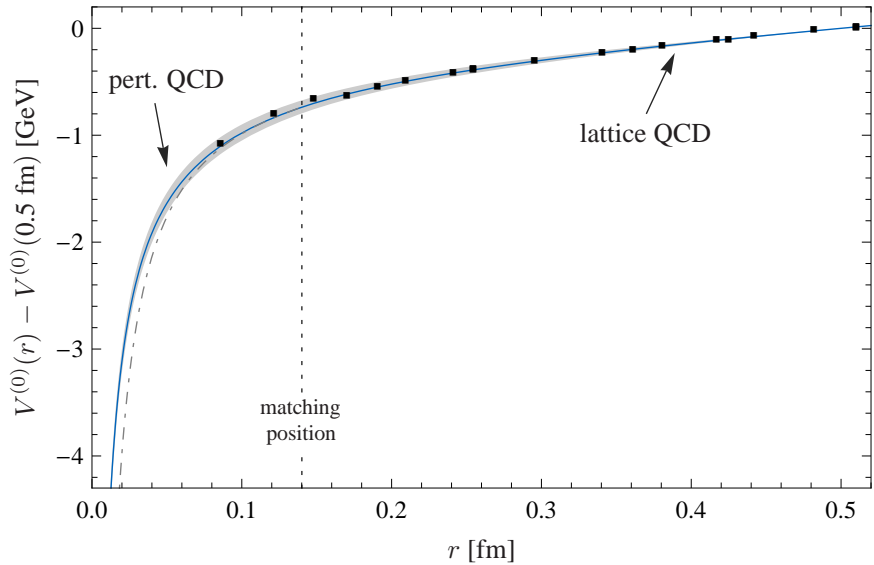}\\
}
\caption{\label{fig:6}Static QCD potential for $n_f=3$, based on
Eqs.~(\ref{eq:r_space_potential}) and~(\ref{eq:static_potential_1}), matched
  at intermediate distances to a potential from lattice QCD~\cite{Bali:2000vr}.
  Dashed-dotted curve: simplest extrapolation using Coulomb-plus-linear
  $r$ dependence.}
\end{figure}
The perturbative potential~(\ref{eq:r_space_potential}), valid at small
distances, can be matched at intermediate distances to results from lattice QCD
(see Fig.~\ref{fig:6}). We use a fit obtained by Bali \emph{et al.}\ from a
full QCD simulation~\cite{Bali:2000vr}. The matching point (dashed line) is
chosen at $r=0.14$~fm. The exact position of the matching point is not important
for the resulting shape of the potential. At $r=0.14$~fm, both the perturbative
and the lattice potential are expected to be reliable. Requiring that the first
derivative of the potential is continuous at the matching point, we find for the
cutoff in Eq.~(\ref{eq:r_space_potential}): $\mu_f=0.908$~GeV (bottomonium
case) and $\mu_f=0.930$~GeV (charmonium case). The grey band reflects
uncertainties in the Sommer scale $r_0= 0.50 \pm 0.03$~fm (lattice part) and
uncertainties in $\alpha_s(|\vec q\,|)$ (perturbative part) as given in the
previous paragraph. This leads to a cutoff window: $\mu_f=0.9^{+0.3}_{-0.2}$~GeV
(for both bottomonium and charmonium). The dashed-dotted line in
Fig.~\ref{fig:6} results from a simple Coulomb-plus-linear extrapolation from
the lattice QCD data to short distances. Evidently, our more sophisticated
perturbative QCD extrapolation based on Eqs.~(\ref{eq:r_space_potential})
and~(\ref{eq:static_potential_1}) differs from that simple form.

\begin{figure}[tp]
{\centering
  \includegraphics[width=.70\textwidth]{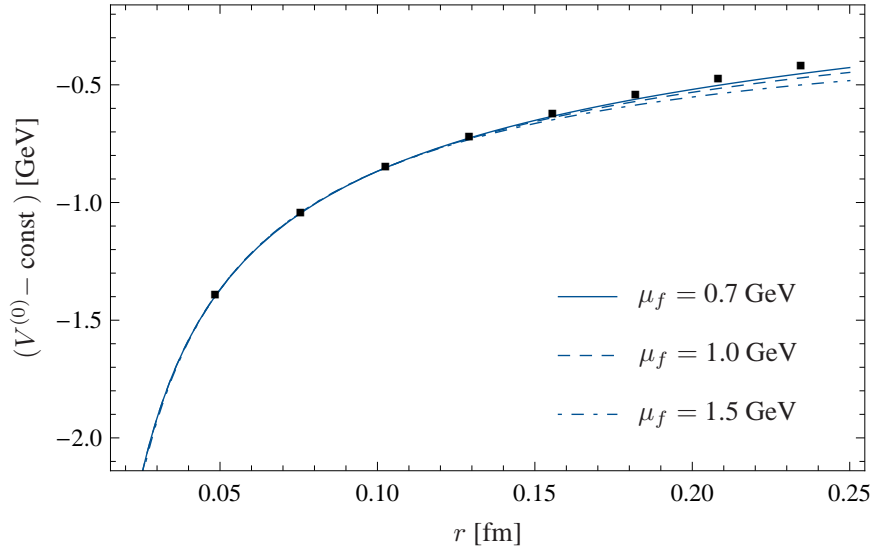}\\
}
\caption{\label{fig:7}The static potential for $n_f\!=\!0$ flavors
compared to lattice points~\cite{Necco:2001xg}. The lattice scale $r_0$ has
been set to $0.5$~fm.}
\end{figure}
For zero flavors one can check against accurate (quenched) lattice
results~\cite{Necco:2001xg} (see Fig.~\ref{fig:7}). Since $\alpha_s$ cannot be
extracted from experiment for $n_f\!=\!0$, we fit to the lattice points below
$0.12$~fm. With a low momentum cutoff $\mu_f$ in the range $0.7$--$1.5$~GeV, we
find $\alpha_s(1.25\ \text{GeV})=0.29\pm0.01$ for the flavorless strong
coupling at the scale of the c-quark mass. The lattice scale $r_0=0.5$~fm has
been used here. A recent precision study of the zero-flavor case in a different
approach can be found in Ref.~\cite{Brambilla:2010pp}.

\section{Order \bm{$1/m$} potential}
\label{sec:order-1/m-potential}
The heavy quark-antiquark potential can be expanded in inverse powers of the
heavy quark mass $m$:
\begin{equation}
  \label{eq:mass_expansion}
  V(r)=V^{(0)}(r)+\frac{V^{(1)}(r)}{m/2}
  +\frac{V^{(2)}(r)}{(m/2)^2}+ \mathcal O (1/m^3)\, .
\end{equation}
The perturbative potential at order $1/m$ in momentum space reads:
\begin{equation}
  \label{eq:q_space_potential_1/m}
  \tilde{V}^{(1)}(|\vec q\,|) = \frac{C_F\,\pi^2\, 
  \alpha_s^2(|\vec q\,|)}{2\,|\vec q\,|}\big\{-C_A
  +\mathcal O(\alpha_s) \big\}\, ,
\end{equation}
with $C_F=4/3$ and $C_A=3$. This form is not unique, see
Ref.~\cite{Brambilla:2000gk}, but we stick to the same convention as the one
used on the lattice. $\tilde{V}^{(1)}(|\vec q\,|)$ can be transformed to
$r$ space as in Eq.~(\ref{eq:r_space_potential}), with a low momentum cutoff
$\mu'_f$ that may differ from $\mu_f$:
\begin{equation}
  \label{eq:r_space_potential_1/m}
  V^{(1)}(r,\mu'_f) = \intop_{|\vec q\,|>\mu'_f}\!\!
  \frac{d^3q}{(2\pi )^3}\ e^{i\vec q\cdot\vec r}\,
  \tilde{V}^{(1)}(|\vec q\,|)\, .
\end{equation}
\begin{figure}[tp]
{\centering
  \includegraphics[width=.70\textwidth]{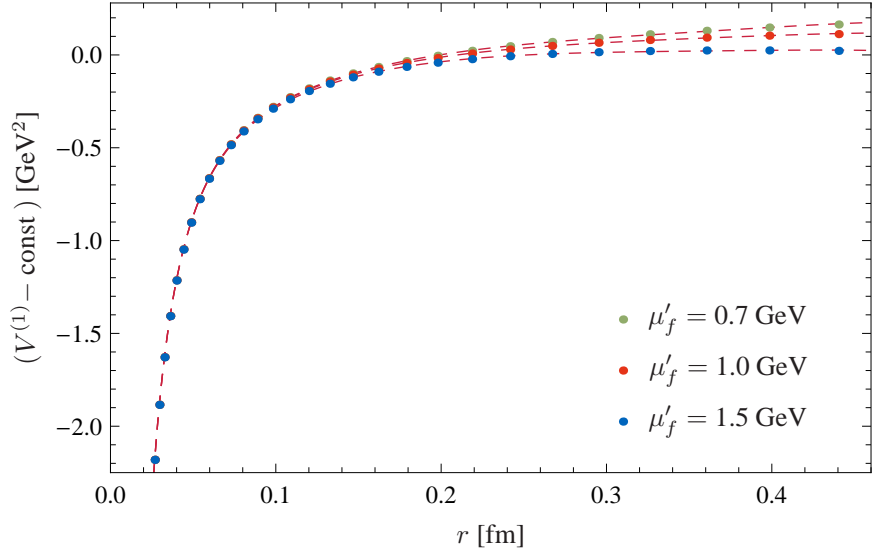}\\
}
\caption{\label{fig:8}The order $1/m$ potential $V^{(1)}(r, \mu_f')$ with
  $n_f=3$ from the restricted numerical Fourier
  transform~(\ref{eq:r_space_potential_1/m}), for different cutoffs $\mu'_f$.
  The curves have been shifted by a constant to match at small $r$ values.}
\end{figure}%
Evidently, the dependence of $V^{(1)}$ on the cutoff scale $\mu'_f$ is again
very weak for  distances $r < 0.2$~fm as shown in Fig.~\ref{fig:8}. The
variation of $(V^{(1)}-\text{const})$ when varying $\mu'_f$ between $0.7$~GeV
and $1.5$~GeV is within $0.02$~GeV$^2$ at the matching radius, $r=0.14$~fm.
This potential is again matched to corresponding results from lattice
QCD~\cite{Koma:2006si,Koma:confinement8}. In order to fit the lattice data we
have used
\begin{equation}
  \label{eq:1/m_lattice_fit}
  V^{(1)}_\text{fit}(r)=-\frac{c'}{r^2}+d' \ln\Big(\frac{r}{r_0}\Big)
  + \text{const}\,,
\end{equation}
with $c'\!=\!0.0027$~GeV$^2\,$fm$^2$, $d'\!=\!0.075$~GeV$^2$ and an arbitrary
length scale $r_0$ that can be absorbed in the overall constant. The
logarithmic form in Eq.~(\ref{eq:1/m_lattice_fit}) is motivated by effective
string theory~\cite{PerezNadal:2008vm}. This parametrization extrapolates the
lattice points for $V^{(1)}$ better than just a $1/r^2\,+\,$linear form. The
lattice calculation of $V^{(1)}$ is quenched and subject to renormalization
issues. A $15\%$ uncertainty is therefore assumed in the lattice potential, in
addition to the uncertainties in the Sommer scale $r_0$.
\begin{figure}[tp]
{\centering
  \includegraphics[width=.70\textwidth]{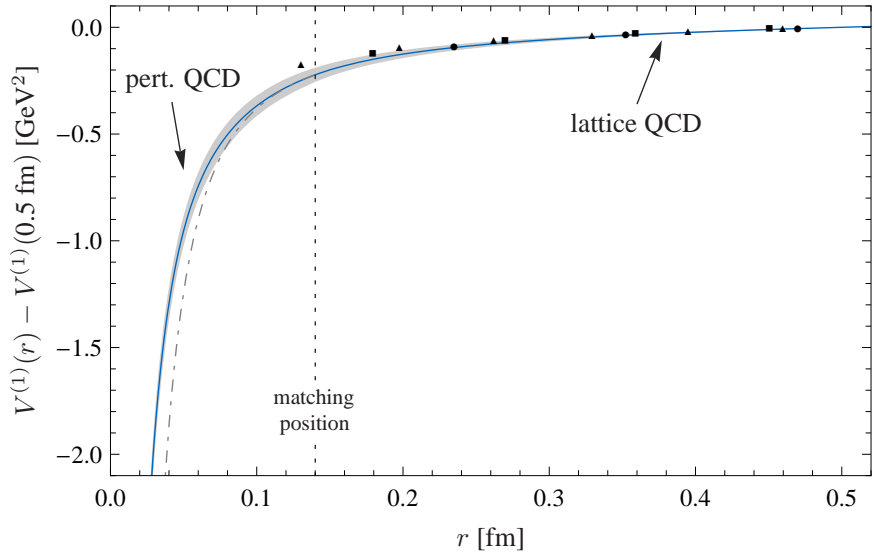}\\
}
\caption{\label{fig:9}The order $1/m$ potential $V^{(1)}(r, \mu_f')$ with
  $n_f=3$, based on Eq.~(\ref{eq:r_space_potential_1/m}), matched at
  intermediate distances to a potential from lattice
  QCD~\cite{Koma:confinement8}. The fit and the matching has been performed
  using the lattice QCD points with $\beta\!=\!5.85$ (circles). Legends are
  analogous to Fig.~\ref{fig:6}. Dashed-dotted curve: simple extrapolation
  using Eq.~(\ref{eq:1/m_lattice_fit}).}
\end{figure}
At short distances a deviation of the perturbative potential $V^{(1)}(r,\mu'_f)$
from $V^{(1)}_\text{fit}$ of Eq.~(\ref{eq:1/m_lattice_fit}) (dashed-dotted line
in Fig.~\ref{fig:9}) is apparent. For the cutoff in
Eq.~(\ref{eq:r_space_potential_1/m}) with error estimate we find:
$\mu'_f=1.9^{+0.4}_{-0.6}$~GeV (bottomonium case) and
$\mu'_f=1.6^{+0.5}_{-0.8}$~GeV (charmonium case).

\section{Quarkonium spectra}
\label{sec:quarkonium-spectra}
Given the potential $V=V^{(0)}+2\,V^{(1)}/m$ up to order $1/m$ in the heavy
quark mass, we can now examine the resulting bottomonium and charmonium
spectra, with focus on the effects of the $1/m$ term. The Schr\"{o}dinger
equation
\begin{equation}
  \label{eq:schroedinger_equation}
  \bigg[-\frac{\hbar^2}{m}\vec \nabla^2+2m_{\widehat{\text{PS}}}(\mu_f,\mu'_f)
  +V^{(0)}(r,\mu_f)+\frac{2}{m}V^{(1)}(r,\mu'_f)-E\bigg]\psi(\vec r\,)=0\, ,
\end{equation}
is solved with the fixed values for $\mu_f$ and $\mu'_f$ as derived in the
construction of the potentials. The potential-subtracted mass
$m_{\widehat{\text{PS}}}(\mu_f,\mu'_f)$, to be defined and discussed in detail
in Section~\ref{sec:quark-masses}, is the only free parameter. It sets the
overall energy scale and it is ultimately fixed by comparison with the measured
bottomonium and charmonium spectrum.

The (heavy) quark mass is not directly measurable. The mass $m$
appearing in the denominator of the kinetic energy and the $V^{(1)}$ term is
therefore not \emph{a priori} determined. In practice we use values close to the
static $\overline{\text{MS}}$ masses: $m\!=\!4.2$~GeV for bottomonium and
$m\!=\!1.25$~GeV for charmonium. Small variations from these values do not have
any significant influence on the results.

\begin{figure}[tp]
{\centering
  \includegraphics[width=.70\textwidth]{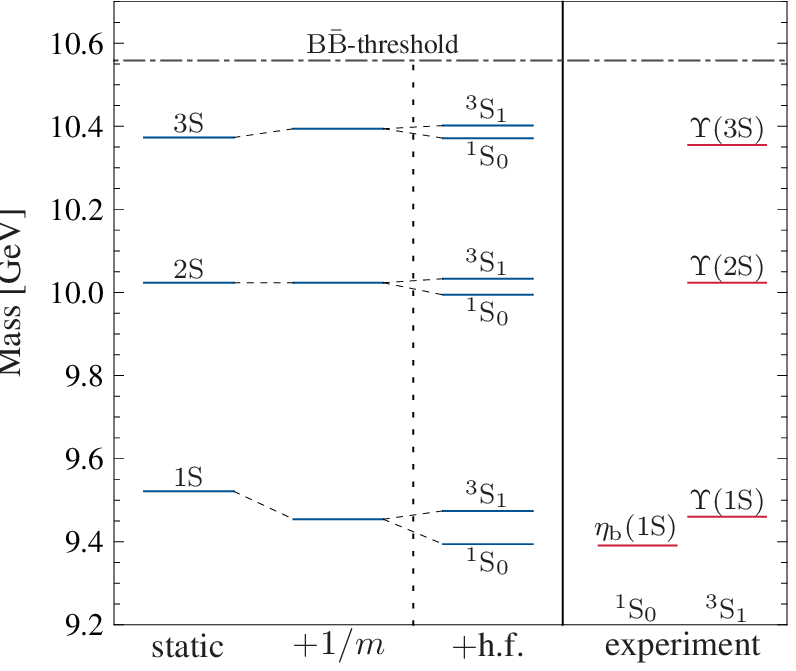}\\
  \includegraphics[width=.70\textwidth]{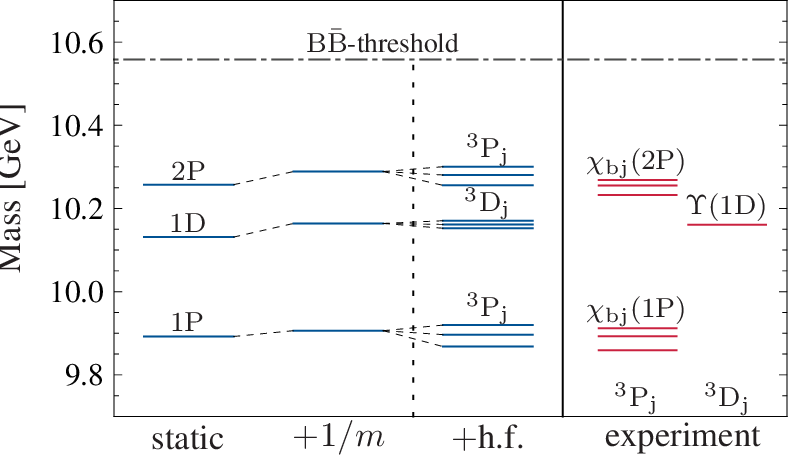}\\
}
\caption{\label{fig:10}Bottomonium spectrum in comparison with experiment.
  Static plus order $1/m$ results are shown on the left, with additional
  hyperfine effects (h.f.) added phenomenologically using
  Eq.~(\ref{eq:effective_one-gluon}).}
\end{figure}
Consider now first the bottomonium spectrum below
$\text{B}\overline{\text{B}}$ threshold\footnote{Above the
$\text{B}\overline{\text{B}}$ threshold the
$\text{b}\overline{\text{b}}$ potential develops an imaginary part and the
present strategy (including lattice QCD) does not apply.}
(see Fig.~\ref{fig:10}). One can start by fixing
$m_{\widehat{\text{PS}}}(\mu_f,\mu'_f)$ such that the measured
$\Upsilon\text{(2S)}$ energy is reproduced. Alternatively, the center of the
$\chi_{\text{b}}\text{(1P)}$ triplet can be used for calibration. These
states remain almost unchanged by the $1/m$ effects. The more tightly bound
$\eta_\text{b}\text{(1S)}$ and $\Upsilon\text{(1S)}$ states respond, as
expected, more sensitively to the non-static corrections induced by
$V^{(1)}(r)$ with its pronounced behavior at short distances.

An additional effective one-gluon exchange spin dependent term,
\begin{equation}
  \label{eq:effective_one-gluon}
  \delta V_{\text{spin}}=
    \frac{8\pi\alpha_s^{\text{eff}}}{9m^2}
    (\vec \sigma_1\!\cdot\!\vec \sigma_2)\,
    \delta^{(3)}(\vec r\,)
    +\frac{\alpha_s^{\text{eff}}}{m^2}\bigg(
    \frac{(\vec \sigma_1\!\cdot\!\vec r\,)
    (\vec \sigma_2\!\cdot\!\vec r\,)}{r^5}\!
    -\!\frac{\vec \sigma_1\!\cdot\!\vec \sigma_2}{3\,r^3}\bigg)
    +\frac{\alpha_s^{\text{eff}}}{m^2}\bigg(
    \frac{\vec L\!\cdot\!\vec \sigma_1
    +\vec L\!\cdot\!\vec \sigma_2}{r^3}\bigg)\, ,
\end{equation}
with $\alpha_s^{\text{eff}}=0.3$ would move all 1S and 1P states well into
their observed positions. For this purpose we replace the delta function (that
is exclusively sensitive to the wave function at $r=0$) by a Gaussian
distribution
\begin{equation}
    \delta^{(3)}(\vec r\,)\to\frac{1}{(\sqrt{\pi}\,\sigma)^3}\, e^{-r^2/\sigma^2},
\end{equation}
with $\sigma=0.03$~fm.
Of course, these procedures based on Eq.~(\ref{eq:effective_one-gluon}) are
purely \emph{ad hoc} and need to be substituted by the full potential of order
$1/m^2$, to be investigated in forthcoming work.

\begin{figure}[tp]
{\centering
  \includegraphics[width=.70\textwidth]{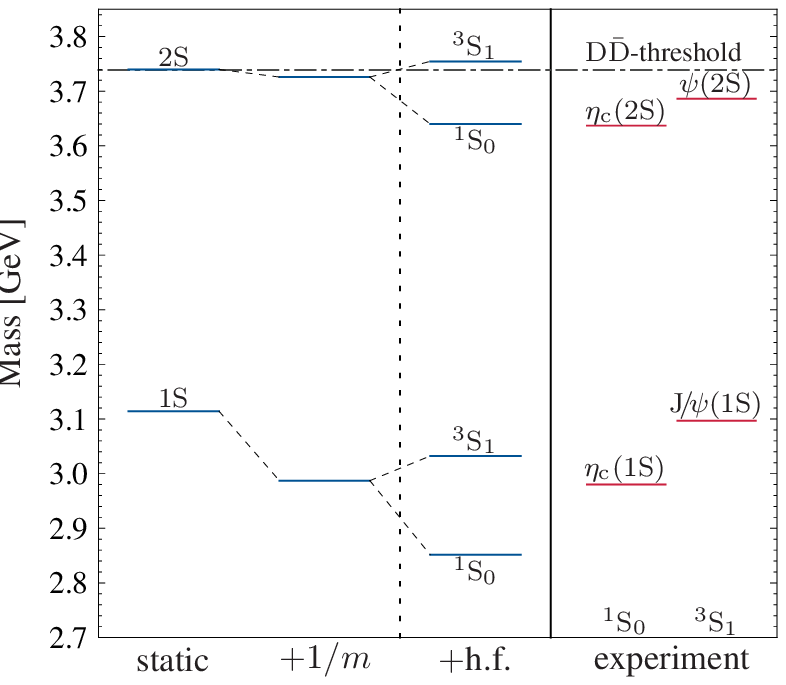}\\
  \includegraphics[width=.70\textwidth]{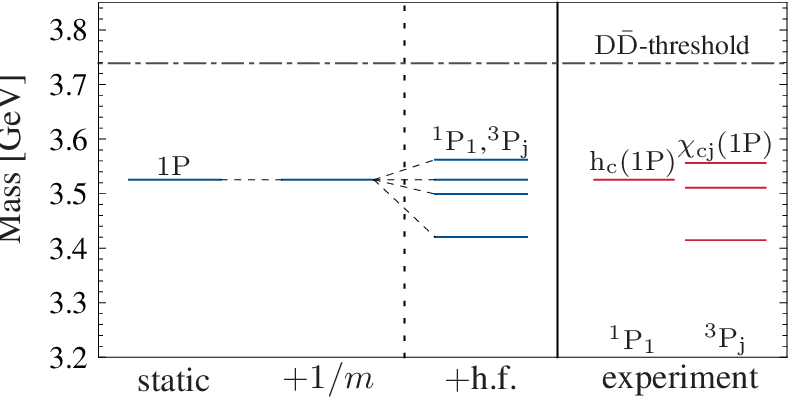}\\
}
\caption{\label{fig:11}Charmonium spectrum in comparison with experiment.
  Static plus order $1/m$ results are shown on the left, with additional
  hyperfine effects (h.f.) added phenomenologically.}
\end{figure}
As expected, the influence of the $1/m$ term in the potential is much stronger
for charmonium than for bottomonium (see Fig.~\ref{fig:11}). For charmonium we
choose the potential-subtracted mass in Eq.~(\ref{eq:schroedinger_equation})
(reflecting the unknown constants in $V^{(0)}$ and $V^{(1)}$) such that the
measured $\text{h}_\text{c}\text{(1P)}$ energy is reproduced without $1/m$
corrections. With this choice, however, the $V^{(1)}$ part produces a downward
shift of $127$~MeV in the 1S states ($\eta_\text{c}$ and $\text{J}\!/\!\psi$)
relative to the static result. This shift is too large in comparison with the
measured $\eta_\text{c}$ and $\text{J}\!/\!\psi$ energies.

The 1S states are naturally more sensitive to $1/m$ corrections than 1P and 2S
states because of the leading $1/r^2$ short-distance behavior of $V^{(1)}$.
Hence the large shift of the 1S energy level at order $1/m$ does not come
unexpectedly. It is nevertheless evident that, no matter which choice is
adopted for adjusting the unknown constant in $V^{(1)}$, the 1S and 1P states
of charmonium cannot be simultaneously reproduced at order $1/m$. Unlike the
situation in bottomonium, corrections of order $1/m^2$ are presumably large in
charmonium.

A manifestation of substantial $1/m^2$ effects is the relatively large observed
splitting of $117$~MeV between $\eta_\text{c}$ and $\text{J}\!/\!\psi$, driven
by an effective coupling strength ($\alpha_s^{\text{eff}}/m^2$ with
$\alpha_s^{\text{eff}}=0.3$ in the phenomenological $\delta V_{\text{spin}}$
of Eq.~(\ref{eq:effective_one-gluon})) that is an order of magnitude larger
than for bottomonium. A systematic investigation of the $1/m^2$ potential,
$V^{(2)}$ in Eq.~(\ref{eq:mass_expansion}), is mandatory now that lattice QCD
data for $V^{(2)}$ are becoming available~\cite{Koma:2006fw,Koma:2010zz}.

\section{Charm and bottom quark masses:\\*
  potential-subtracted and $\overline{\text{MS}}$ schemes}
\label{sec:quark-masses}
The static potential is determined up to an overall constant. Introducing the
potential-subtracted (PS) quark mass as
\begin{equation}
  \label{eq:ps_mass}
  m_\text{PS}(\mu_f)=m_\text{pole} + \frac{1}{2}\intop_{|\vec q\,|<\mu_f}\!
  \frac{d^3 q}{(2\pi )^3}\,\tilde{V}^{(0)}(|\vec q\,|)
  =m_\text{pole} + \frac{1}{4\pi^2}\intop_0^{\mu_f}\!
  dq\,q^2\,\tilde{V}^{(0)}(q)
\end{equation}
in terms of the pole mass of the charm or bottom quark, this $m_\text{PS}$
absorbs the unknown constant and does not suffer from the leading renormalon
ambiguity~\cite{Beneke:1998rk,Hoang:1998nz}. With the previous choice of the
constant to reproduce the measured $\Upsilon\text{(2S)}$ energy, we find the
value $m_\text{PS}(\mu_f\!=\!0.908\ \text{GeV})=4.78$~GeV in the PS mass scheme.
In the charmonium case we fit to the $\text{h}_\text{c}(\text{1P})$ energy and
find $m_\text{PS}(\mu_f\!=\!0.930\ \text{GeV})=1.39$~GeV. To convert the PS
mass to the mass in the more commonly used $\overline{\text{MS}}$ scheme, it is
necessary to introduce the pole mass $m_\text{pole}$ as an intermediate step.
The relation between $m_\text{pole}$ and $m_\text{PS}$
reads~\cite{Beneke:1998rk}:
\begin{multline}
  \label{eq:mass_relation_1}
  m_\text{pole} = m_\text{PS}(\mu_f)+\frac{C_F\, \alpha_s(\mu)\,\mu_f}{\pi}
  \bigg\{1+\frac{\alpha_s(\mu)}{4\pi}
  \bigg[a_1-\beta_0 \bigg(\!\ln\frac{\mu_f^2}{\mu^2}-2\bigg) \bigg]\\
  +\bigg(\frac{\alpha_s(\mu)}{4\pi}\bigg)^2
  \bigg[a_2-(2a_1\beta_0\!+\!\beta_1)\bigg(\!\ln\frac{\mu_f^2}{\mu^2}
  -2\bigg)+\beta_0^2\bigg(\!\ln^2\frac{\mu_f^2}{\mu^2}
  -4\ln\frac{\mu_f^2}{\mu^2}+8\bigg) \bigg]+\mathcal O(\alpha_s^3) \bigg\}\, ,
\end{multline}
with the same conventions as in Section~\ref{sec:first-attempts}. Note that a
renormalization scale $\mu$ appears in the coupling. In the following $\mu$ is
set equal to the $\overline{\text{MS}}$ mass
$\overline m\equiv m_{\overline{\text{MS}}}(m_{\overline{\text{MS}}})$. This
$\overline m$ is not known at that point and has to be computed iteratively.
In a second step the pole mass is converted to the $\overline{\text{MS}}$
mass~\cite{Chetyrkin:1999qi,Melnikov:2000qh}:
\begin{align}
  \label{eq:mass_relation_2}
  \frac{m_\text{pole}}{\overline m} = 1&+\frac{4}{3}
  \bigg(\frac{\alpha_s(\overline m)}{\pi}\bigg)
  +\bigg(\frac{\alpha_s(\overline m)}{\pi}\bigg)^2(-1.0414\,n_f+13.4434)\notag\\
  &+\bigg(\frac{\alpha_s(\overline m)}{\pi}\bigg)^3
  (0.6527\,n_f^2-26.655\,n_f+190.595)+\mathcal O(\alpha_s^4)\, .
\end{align}
Note that both relations (\ref{eq:mass_relation_1}) and
(\ref{eq:mass_relation_2}), taken individually, show a poorly convergent
behavior whereas the relation between $m_\text{PS}(\mu_f)$ and $\overline m$
is expected to be stable. This is in fact confirmed numerically. Using
$\mu_f=0.908$~GeV (from Section~\ref{sec:static-potential}) one obtains the
value $\overline m_\text{b}=4.27$~GeV for the bottom quark;
with $\mu_f=0.930$~GeV one finds $\overline m_\text{c}=1.24$~GeV for the charm
quark.

\begin{figure}[tp]
{\centering
  \includegraphics[width=.70\textwidth]{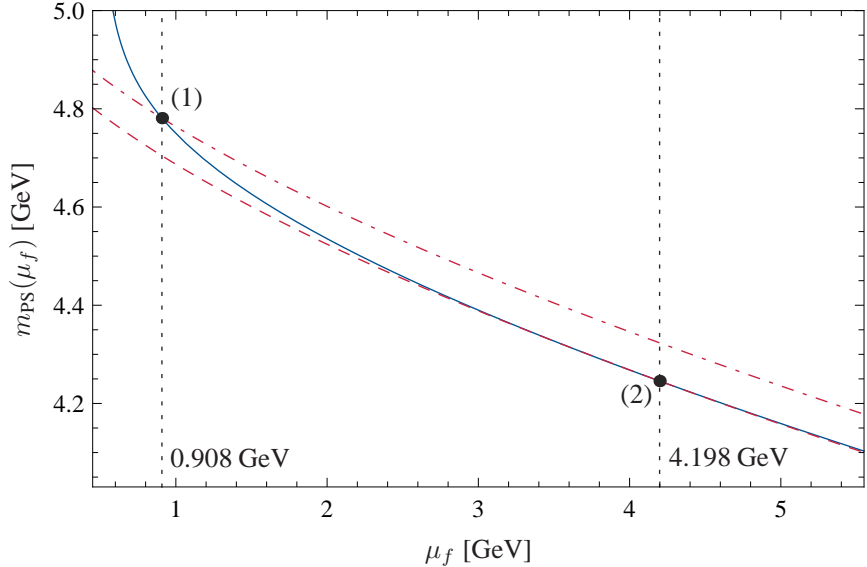}\\
}
\caption{\label{fig:12}Scale dependence of $m_{\text{PS}}(\mu_f)$ for the
  bottomonium case. The solid curve shows the numerical $\mu_f$ dependence from
  Eq.~(\ref{eq:ps_mass}). The dashed and dashed-dotted curves show the $\mu_f$
  dependence given by Eq.~(\ref{eq:mass_relation_1}) at different matching
  points. Point (2) is preferred to point (1) for the determination of
  $\overline m_\text{b}$.}
\end{figure}
However, as shown in Fig.~\ref{fig:12}, the $\mu_f$ dependence of
$m_\text{PS}(\mu_f)$ (Eq.~(\ref{eq:mass_relation_1})) differs for
$\mu_f\!\ll\!\overline m$ from the $\mu_f$ dependence coming from variations of
the cutoff in the numerical integral (solid line). Instead of matching at
$\mu_f=0.908$~GeV it is obviously preferable to use the numerical $\mu_f$
dependence first to translate $m_\text{PS}(\mu_f)$ into
$m_\text{PS}(\overline m)$ and then apply Eqs.~(\ref{eq:mass_relation_1})
and~(\ref{eq:mass_relation_2}) to translate this value into the
$\overline{\text{MS}}$ scheme. By this method the extraction of $\overline m$
becomes independent of the value of $\mu_f$ used in the construction of the
potential. This leads to improved mass values, $\overline m_\text{b}=4.20$~GeV
for the bottom quark and $\overline m_\text{c}=1.23$~GeV for the charm quark in
the $\overline{\text{MS}}$ scheme.

To determine values for the quark masses at order $1/m$ a redefinition of the
PS mass is required:
\begin{equation}
  m_{\widehat{\text{PS}}}(\mu_f,\mu_f') \equiv m_{\text{PS}}(\mu_f)
  -\frac{1}{8m}C_F C_A  \alpha_s^2(\overline m)\,\mu_f'^{\,2}\, .
\end{equation}
The $1/m$-term stems from an analogous calculation as in the static case. The
renormalization scale $\mu$ that appears in the coupling has again been
identified with $\overline m$. We determine
$m_{\widehat{\text{PS}}}(\mu_f,\mu_f')$ for the b- and c-quark by fitting to the
empirical $\Upsilon\text{(2S)}$ and $\text{h}_\text{c}(\text{1P})$ energies,
respectively, and convert these values numerically to
$m_{\widehat{\text{PS}}}(\overline m,\overline m)$. This leads in a second step
to the $1/m$-improved $\overline{\text{MS}}$ values
$\overline m_\text{b}=4.18$~GeV for the bottom quark and
$\overline m_\text{c}=1.28$~GeV for the charm quark. In Table~\ref{tab:t1} the
quark masses in the $\overline{\text{MS}}$ scheme found in our approach are
summarized and compared to values given by the Particle Data
Group~\cite{Nakamura:2010zzi}. 
\begin{table}[tp]
{\centering
  \begin{tabular}{|l|c|c||c|}
  \hline
  \multicolumn{4}{|c|}{ $\overline{\text{MS}}$ masses [GeV]}\\
  \hline
  &Static&Static + $\mathcal O (1/m)$&PDG 2010\\
  \hline
  Bottom quark&$4.20\pm 0.04$
    &$4.18^{+0.05}_{-0.04}$
    &$4.19^{+0.18}_{-0.06}\vphantom{\Big(}$\\
  Charm quark&$1.23\pm 0.04$
    &$1.28^{+0.07}_{-0.06}$
    &$1.27^{+0.07}_{-0.09}\vphantom{\Big(}$\\
  \hline
  \end{tabular}\\
}
\caption{\label{tab:t1}Comparison of quark masses obtained in our approach
(leading order plus order~$1/m$ corrections) with the values listed by the
Particle Data Group (PDG)~\cite{Nakamura:2010zzi}. See text for details
concerning error estimates.}
\end{table}
We have performed error estimates for the quark masses, reflecting
uncertainties in the potentials (static and order $1/m$). Additional
uncertainties are included from our specific choice of matching to the
empirical $\Upsilon\text{(2S)}$ and $\text{h}_\text{c}(\text{1P})$ energies.
The errors at order $1/m$ have increased in comparison to those for the static
case since they incorporate in addition the error band from $V^{(1)}$. The
error estimates at order $1/m$ do not include possible further uncertainties
appearing at order $1/m^2$.

\section{Summary}
Improved bottomonium and charmonium potentials have been derived up to and
including order $1/m$ in the heavy quark masses by systematically matching
perturbative QCD results to accurate lattice QCD data at an intermediate
distance scale, $r=0.14$~fm. A single constant (the potential-subtracted quark
mass) is adjusted to reproduce the $\Upsilon\text{(2S)}$ and
$\text{h}_\text{c}(\text{1P})$ masses, respectively. The predicted pattern of
all other subthreshold states agrees well with the empirical bottomonium
spectroscopy. For charmonium, $1/m$ effects are far more pronounced, as
expected. The potential-subtracted heavy quark masses at order $1/m$, when
translated to $\overline{\text{MS}}$ masses, agree well with those listed by
the Particle Data Group. In a next step, corrections of order $1/m^2$ will be
studied along the same lines.

\begin{acknowledgments}
This work was supported in part by BMBF, GSI and the DFG Excellence Cluster
``Origin and Structure of the Universe''. We thank Antonio Vairo and Nora
Brambilla for numerous useful discussions and helpful comments. We are also
grateful to Hartmut Wittig for instructions concerning lattice QCD results.
One of the authors (A.\,L.) acknowledges the support of the TUM Graduate School.
\end{acknowledgments}

\end{document}